\documentclass[twocolumn]{jpsj3}

\usepackage{txfonts}
\usepackage{footmisc}
\usepackage{comment}
\usepackage{color}  % This corrects the distorted eps figures

\title{DFT-Based Engineering of Dirac Surface Energy \\ in Topological-Insulator Multilayers}

\author{Takao Kosaka, Kunihiko Yamauchi\thanks{kunihiko@sanken.osaka-u.ac.jp} and Tamio Oguchi}
\inst{Institute of Scientific and Industrial Research, Osaka University, 8-1 Mihogaoka, Ibaraki, Osaka 567-0047, Japan} %\\

\abst{
Aiming at the future spintronics device applications of the spin-polarized surface states in three-dimensional topological insulator, 
a highly insulating bulk state and a tunable Dirac cone surface state are required.   
Here we employ a slab model having hetero-structural Bi$_{2}$Se$_{3}$-related quintuple layers 
and perform first-principles simulations.  
 Our computational results show that the Dirac-point energy can be optimally tuned by selecting an appropriate pair of materials 
 so that the work function at the surface quintuple layer is slightly different from that at the inner quintuple layers. 
 The ideal surface state is obtained in Bi$_{2}$Te$_{3}$/(Bi$_{2}$Te$_{2}$Se)$_{4}$/Bi$_{2}$Te$_{3}$ slab, in which the Fermi lines show the significant warping effect 
 and  both the in-plane and the out-of-plane components of the spin polarization emerge.  
}

\newcommand{\bs}{Bi$_{2}$Se$_{3}$}
\newcommand{\bt}{Bi$_{2}$Te$_{3}$}
\newcommand{\bts}{Bi$_{2}$Te$_{2}$Se} 
\newcommand{\st} {Sb$_{2}$Te$_{3}$}
\newcommand{\sbs} {Sb$_{2}$Se$_{3}$}

\newcommand{\btbts}{$\mathrm{Bi_{2}Te_{3}}$/$(\mathrm{Bi_{2}Te_{2}Se})_{4}$/$\mathrm{Bi_{2}Te_{3}}$}
\newcommand{\sbbts}{$\mathrm{Sb_{2}Te_{3}}$/$(\mathrm{Bi_{2}Te_{2}Se})_{4}$/$\mathrm{Sb_{2}Te_{3}}$}

\begin{document}
\maketitle

\section{Introduction}
%Topological insulators exhibit the non-trivial quantum states which can be characterized by the insulating bulk states and spin-polarized metallic surface states,\cite{topo_1}\cite{topo_2}\cite{topo_3}\cite{topo_4} such as surface states with Dirac cone, being expected for spintronics device applications. 
A three-dimensional topological insulator (TI) exhibits the non-trivial quantum states which can be characterized by the insulating bulk states and the spin-polarized metallic surface states\cite{topo_1, topo_2, topo_3, topo_4}.
The latter is known to have the Dirac cone dispersion and show the  spin polarized state, 
which has an appealing potential for future spintronics device applications.
%The highly insulating bulk state and the tunable Dirac cone are required for the future device applications. 
%Nevertheless, many of topological insulators have been found to be metallic due to the existence of impurities and disorder. 
Although the highly insulating bulk state and the tunable Dirac cone in the bulk band gap are desired for the   applications, 
 many of topological insulators have been found to be metallic due to the existence of impurities and/or structural disorder. 
For example, in Bi$_{2}$Se$_{3}$, Se vacancies and Se anti-site defects lead to  electron carrier doping in the conduction bands.\cite{defect_1, defect_2, defect_BS_1, defect_BS_2, defect_BS_3} 
In Bi$_{2}$Te$_{3}$, Te (Bi) anti-site defects leads to electron (hole) carrier doping.\cite{defect_1}\cite{defect_2}\cite{defect_BT}  
It has been also observed that the Dirac point is embedded in the bulk valence bands in Bi$_{2}$Te$_{3}$\cite{bi2te3.arpes, BT_1, BT_2}, which makes impossible to utilize the Dirac carrier. 
%In fact, Bi$_{2}$Se$_{3}$ crystals are found to be electron doped and become n-type,\cite{defect_BS_1}\cite{defect_BS_2}\cite{defect_BS_3} and Bi$_{2}$Te$_{3}$ crystals can be found to be both n- and p- types.\cite{defect_BT} 
%It has been reported that the ternary tetradymite topological insulator materials such as $\mathrm{Bi_{2}Te_{2}Se}$\cite{BST_1}\cite{BST_2} which forms the ordered Te-Bi-Se-Bi-Te quintuple layers and $\mathrm{Bi_{2-x}Sb_{x}Te_{3-y}Se_{y}}$ solid solution\cite{BSTS_1}\cite{BSTS_2}\cite{BSTS_3} have highly bulk insulating states, because the defect formation is suppressed in these materials. 
%In addition to such defect control, one can obtain spin-polarized Dirac carries with the Dirac point in the band gap. 
Recently, it has been reported that %the ternary tetradymite topological insulator materials such as 
$\mathrm{Bi_{2}Te_{2}Se}$\cite{BST_1, BST_2} %which forms the ordered Te-Bi-Se-Bi-Te quintuple layers 
and $\mathrm{Bi_{2-x}Sb_{x}Te_{3-y}Se_{y}}$ solid solution\cite{BSTS_1, BSTS_2, BSTS_3} show highly bulk insulating states, because the defect formation can be suppressed by controlling the solid solution ratio in these materials. 
In these materials, the spin-polarized Dirac cones appear in the middle of the band gap, and therefore 
they are now considered as a good playground to experimentally investigate the topological surface bands. 
In the present study, by using first-principles calculations, 
we propose an alternative way  to tune the Dirac-cone energy, that is, 
artificial stacking control of the  Bi$_{2}$Se$_{3}$-related  quintuple-layer slabs.

%The Dirac surface states of Bi$_{2}$Se$_{3}$ is simple and almost ideal structure.\cite{topo_3} 
%On the other hand, the Dirac surface states of Bi$_{2}$Te$_{3}$ is complicated structure.\cite{BT_1}\cite{BT_2}  
%The Dirac point of the surface stats locates in the bulk valence band. 
%This makes it difficult to probe the surface transport properties near the Dirac point without being disturbed by bulk carriers in Bi$_{2}$Te$_{3}$. 
%As a result, the Dirac carrier can't be used in Bi$_{2}$Te$_{3}$. 
%In addition, the constant-energy contour of the Dirac cone in Bi$_{2}$Te$_{3}$ are hexagonal warping structure and the out-of-plane spin component on the surface states appear. \cite{sz}
%This warping and out-of-plane spin component are caused by spin-orbit coupling at the surface of rhombohedral crystal systems.\cite{warp} 
%In this paper, we control the Dirac surface states by engineering topological multilayer structures using first-principles calculations based on density functional theory. 

\section{Method}

\subsection{Structural Models}

\begin{figure}
\center
\includegraphics[width=7cm]{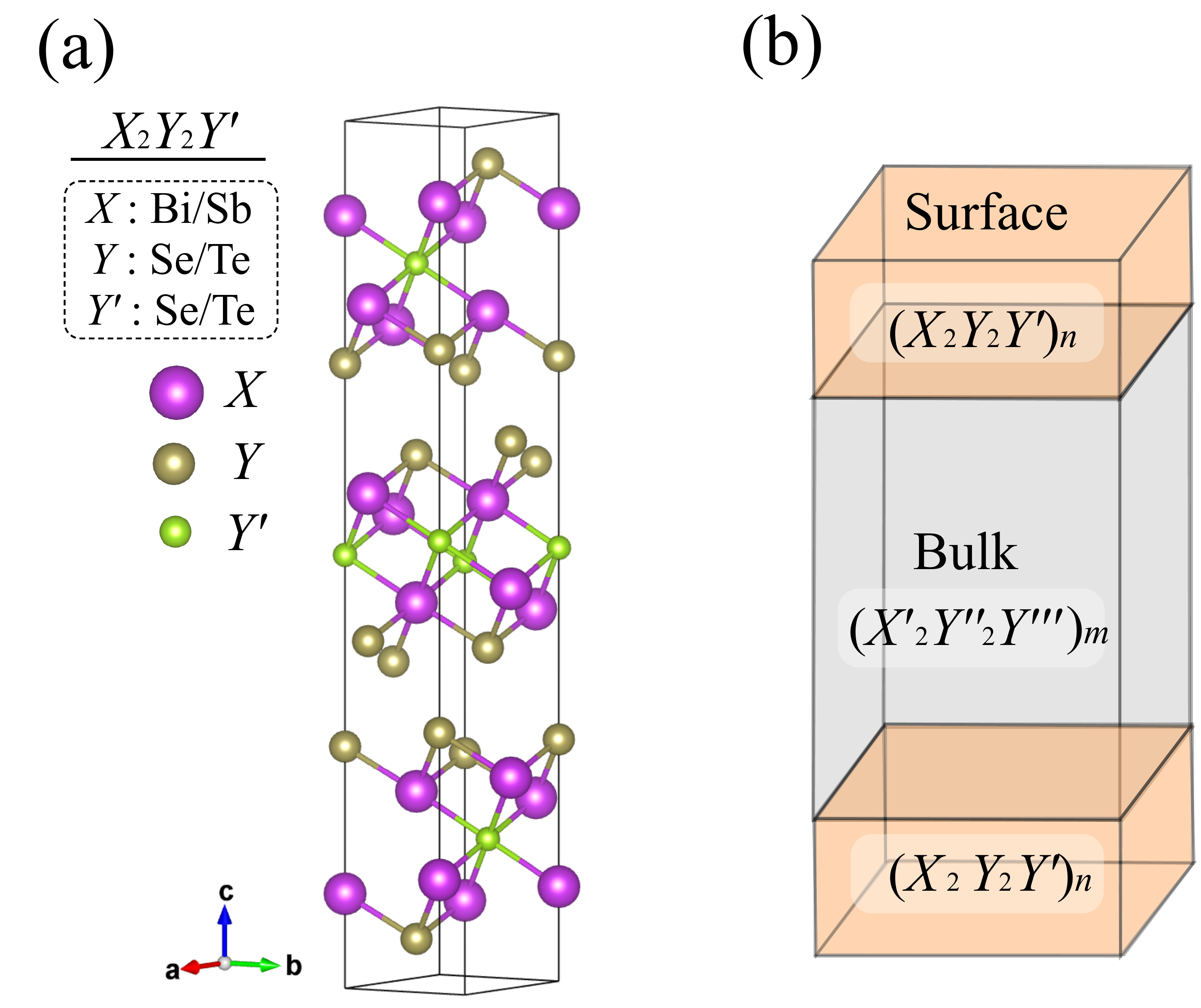}
\caption{(a) Crystal structure of tetradymite chalcogenides $X_{2}Y_{2}Y'$, such as $\mathrm{Bi_{2}Se_{3}}$ ($\it{X}$ = Bi, $\it{Y}$ = $\it{Y'}$ = Se), $\mathrm{Bi_{2}Te_{3}}$ ($\it{X}$ = Bi, $\it{Y}$ = $\it{Y'}$ = Te), $\mathrm{Bi_{2}Te_{2}Se}$ ($\it{X}$ = Bi, $\it{Y}$ = Te, $\it{Y'}$ = Se), $\mathrm{Sb_{2}Te_{3}}$ ($\it{X}$ = Sb, $\it{Y}$ = $\it{Y'}$ = Te) and $\mathrm{Sb_{2}Se_{3}}$ ($\it{X}$ = Sb, $\it{Y}$ = $\it{Y'}$ = Se), in a hexagonal setting. 
%Multilayers consist of two kinds of topological insulator materials. 
(b) A schematic structure of the hetero-structural multi-QL slab. % ($n+m+n=6$). %See  main text for the details. 
%The structure of multilayers is $\it{n}$-$\mathrm{\it{X}_{\mathrm{2}}\it{Y}_{\mathrm{2}}\it{Y'}}$/$\it{m}$-$\mathrm{\it{X'}_{\mathrm{2}}\it{Y''}_{\mathrm{2}}\it{Y'''}}$/$\it{n}$-$\mathrm{\it{X}_{\mathrm{2}}\it{Y}_{\mathrm{2}}\it{Y'}}$ ($\it{X}$, $\it{X'}$ = Bi or Sb, and $\it{Y}$, $\it{Y'}$, $\it{Y''}$, $\it{Y'''}$ = Te or Se). 
%This structure remains inversion symmetry.
}
\label{fig:st}
\end{figure}

%$\mathrm{Bi_{2}Se_{3}}$, $\mathrm{Bi_{2}Te_{3}}$, $\mathrm{Bi_{2}Te_{2}Se}$ and $\mathrm{Sb_{2}Te_{3}}$ are three dimensional topological insulator. 
%These form the ternary tetradymite structure (Fig. \ref{fig:st} (a)). 
%They consists of covalently bonded quintuple layers ( $\it{Y}$-$\it{X}$-$\it{Y'}$-$\it{X}$-$\it{Y}$ ). 
%Each quintuple layers are weakly connected by van der Waals interaction. 
%The stack of quintuple layers are in the -A-B-C-A-B-C- manner.
%\par
%Topological insulator multilayers consist of two kinds of topological insulator materials. 
Three dimensional topological insulators, $\mathrm{Bi_{2}Se_{3}}$, $\mathrm{Bi_{2}Te_{3}}$, $\mathrm{Bi_{2}Te_{2}Se}$ and $\mathrm{Sb_{2}Te_{3}}$, crystalize in  
ternary tetradymite structure as shown in Fig. \ref{fig:st} (a). 
The covalently bonded quintuple layers (QLs) %: $\it{Y}$-$\it{X}$-$\it{Y'}$-$\it{X}$-$\it{Y}$)  
 are weakly stacked in the ABCABC sequence  through van der Waals interaction. 
%\par
In this study, we set up a multi-QL periodic slab model which contains 
two kinds of topological-insulator materials selected out of 
$\mathrm{Bi_{2}Se_{3}}$, $\mathrm{Bi_{2}Te_{3}}$, $\mathrm{Bi_{2}Te_{2}Se}$, $\mathrm{Sb_{2}Te_{3}}$ and $\mathrm{Sb_{2}Se_{3}}$ (Note: among them, only $\mathrm{Sb_{2}Se_{3}}$ is a trivial insulator). 
%We also checked other slabs that consists of larger number of QLs.  
The multi-QL sequence is described as $\mathrm{(\it{X}_{\mathrm{2}}\it{Y}_{\mathrm{3}})_{n}}$/$\mathrm{(\it{X}_{\mathrm{2}}\it{Y}_{\mathrm{3}})_{m}}$/$\mathrm{(\it{X}_{\mathrm{2}}\it{Y}_{\mathrm{3}})_{n}}$ ($\it{X}$ = Bi or Sb, and $\it{Y}$ = Te or Se) as shown in Fig. \ref{fig:st} (b). For the sake of simplicity, we consider only the centrosymmetric slab structure. We also validated that the six QL slab is thick enough to obtain the topologically protected surface bands.

\subsection{Computational Details}

Bandstructure calculations were performed by using a projector augmented wave method\cite{PAW} implemented in Vienna Ab initio Simulation Package (VASP) code\cite{VASP} with generalized gradient approximation (GGA)\cite{PBE} to the density functional theory (DFT). 
To obtain accurate inter-QL distances, long-range van der Waals interaction was taken into account through semi-empirical corrections by DFT-D2 approach.\cite{D2} 
After the crystal structure was fully optimized until forces acting on atoms were smaller than $1 \times 10^{-5}$ eV/\AA, the spin-orbit coupling was included self-consistently. 
The $\it{k}$-point mesh was set to be $12 \times 12 \times 1$. 
The in-plane experimental lattice parameter was used for the calculation;  
for \bs, \bts, \bt, \sbs, and \st, 
$a$=4.143 \cite{lat.bi2se3}, 4.280 \cite{lat.bi2te2se}, 4.382 \cite{lat.bi2te3}, 4.034 \cite{lat.sb2se3}, 4.250 \cite{lat.sb2te3} \AA\ were used,
respectively. 
For the hetero-structural slab calculation, the lattice parameter of the material located at the inner (bulk) part of slab was employed. 
The vacuum layer was set to be larger than 25 \AA\ between the slabs.

%\begin{figure*}
%\center
%\includegraphics[width=17cm]{band_o.pdf}
\begin{figure}[htbp]
\center
\includegraphics[width=6.7cm]{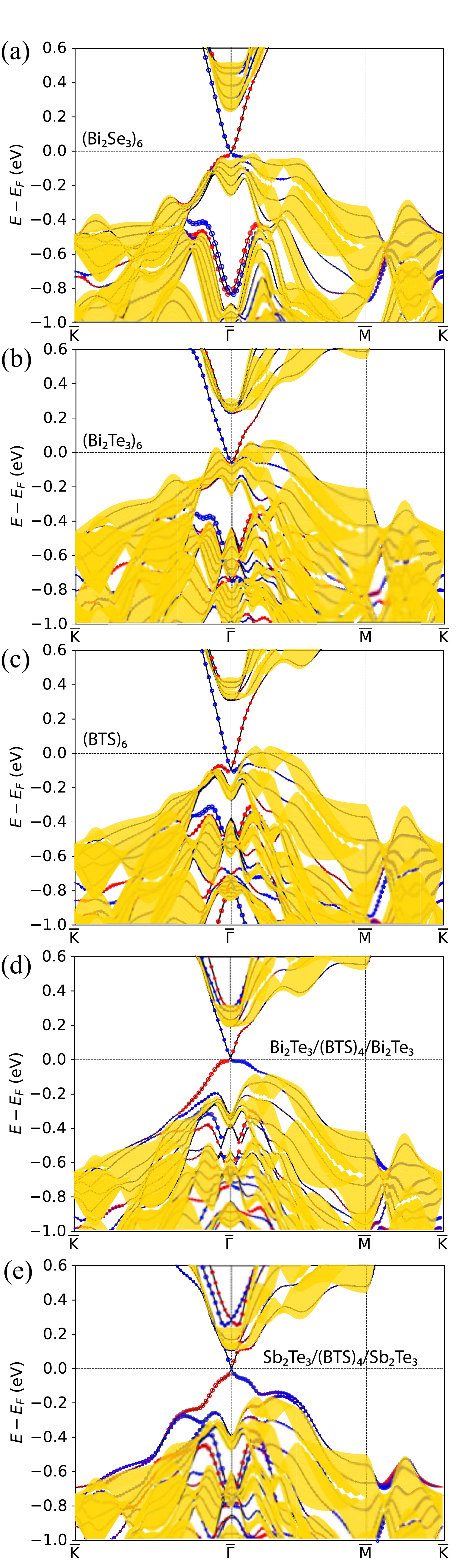}
\caption{
Calculated bandstructure of (a) (\bs)$_{6}$, (b) (\bt)$_{6}$, (c) (\bts)$_{6}$
(d) \btbts\ and (e) \sbbts\
 slabs. 
%(a) Schematic band structure of Bi$_{2}$Te$_{2}$Se base multilayers. 
%Multilayer structure is $\it{X}_{\rm{2}}$$\it{Y}_{\rm{2}}$$\it{Y'}$/4-Bi$_{2}$Te$_{2}$Se/$\it{X}_{\rm{2}}$$\it{Y}_{\rm{2}}$$\it{Y'}$. 
%Surface layer is $\it{X}_{\rm{2}}$$\it{Y}_{\rm{2}}$$\it{Y'}$. 
%(b), (c), (d) Band structure of Bi$_{2}$Te$_{2}$Se base multilayers. 
%Band structure of a symmetric six quintuple layers slab (black lines). 
Red and blue circles indicate the in-plane up and down spin polarization respectively, while 
the circle size is proportional to the weight contributed from the outermost surface QL. 
Bulk band projections (bandstructure of bulk crystal, summation of ten slices along $k_{z}$) are shown by the yellow shade areas. 
%(b) Band structure of a symmetric six quintuple layers $\mathrm{Bi_{2}Te_{2}Se}$ slab. 
%(c) $\mathrm{Bi_{2}Te_{3}}$/4-$\mathrm{Bi_{2}Te_{2}Se}$/$\mathrm{Bi_{2}Te_{3}}$ multilayers slab. 
%The slab consist of six quintuple layers. 
%Surface quintuple layer is $\mathrm{Bi_{2}Te_{3}}$, and bulk four quintuple layers are made of $\mathrm{Bi_{2}Te_{2}Se}$. 
%(d) $\mathrm{Sb_{2}Te_{3}}$/4-$\mathrm{Bi_{2}Te_{2}Se}$/$\mathrm{Sb_{2}Te_{3}}$ multilayers slab. 
}
\label{fig:band5}
\end{figure}

\section{Results}
\subsection{Homo-structural TI slabs}

\begin{figure}[htbp]
\center
\includegraphics[width=7cm]{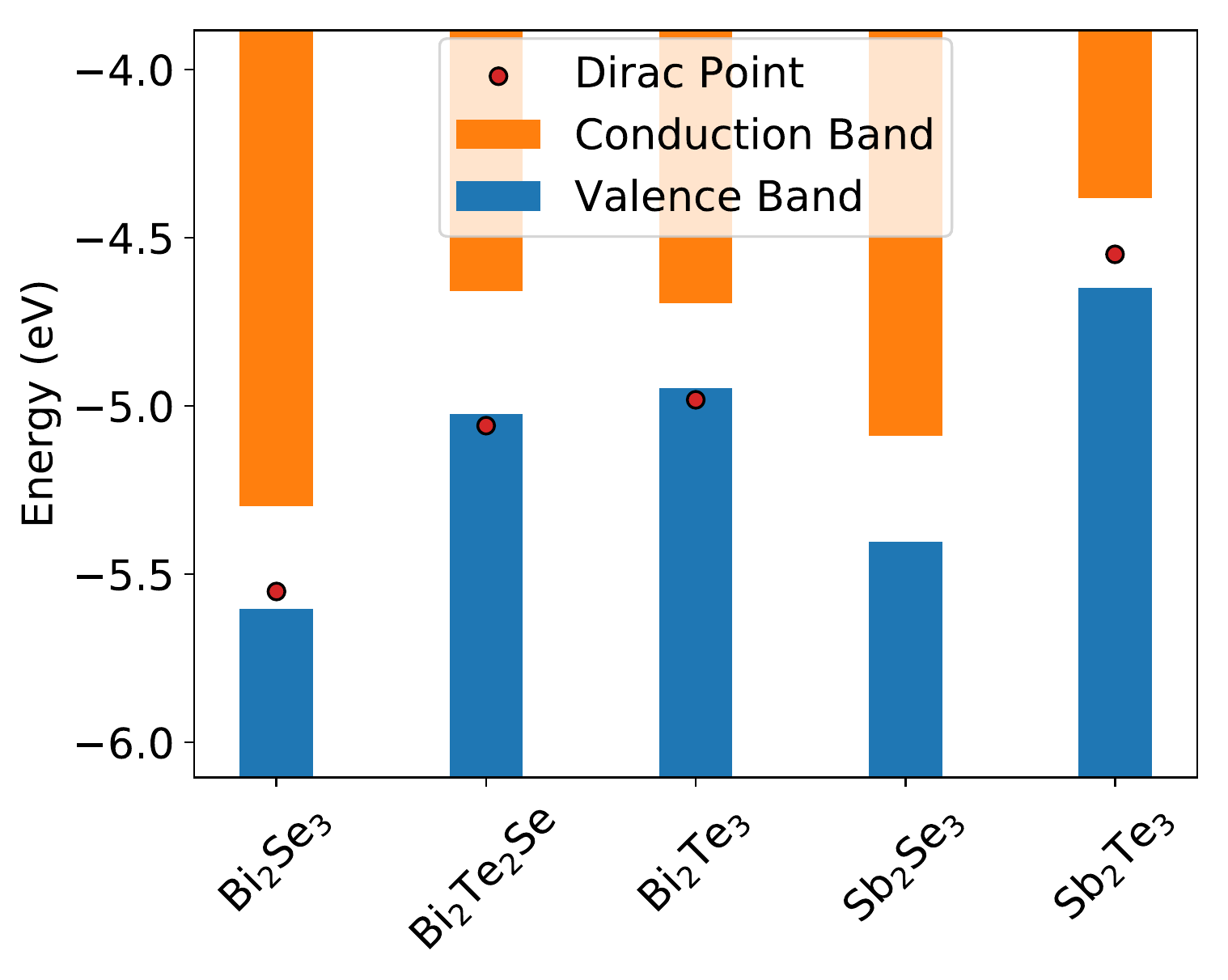}
\caption{Energy diagram of the bulk valence band maximum and bulk conduction band minimum with respect to the vacuum level,  % (0eV is set to vacuum level $E_{\rm{Vacuum \: Level}}$). 
 calculated in six QL slab. 
%In this paper, the bulk band is regarded as the contribution of the bulk layers to the band structure. 
%The bulk layers are inner quintuples layers which exclude the outermost surface quintuple layer.
}
\label{fig:cbvb}
\end{figure}

First we look at the calculated bandstructures of homo-structural slabs, i.e., 
(\bs)$_{6}$, (\bt)$_{6}$, and (\bts)$_{6}$ slabs,
 as shown in Fig. \ref{fig:band5} (a)-(c).  %together with the bulk bands projected onto ($kx, ky$) planes. 
%As summarized in Tbl.\ref{t1}, 
The band gap of  (\bs)$_{6}$ is calculated as $E_{\rm gap}$=0.31 eV, which is consistent with the experimentally measured band gap of bulk \bs, $E_{\rm gap}^{\rm exp}$=0.2-0.3 eV.\cite{expgap1, expgap2} 
(\bt)$_{6}$ shows smaller band gap, $E_{\rm gap}$=0.27 eV. 
Among the three slabs, the band gap is largest, $E_{\rm gap}=0.37$eV, at (\bts)$_{6}$. 
In the slab calculation, the inner four QLs forms the bulk-like bands whereas an outmost QL forms the surface bands. 
The valence band maximum (VBM) of the bulk QL  is found along the $\mathrm{\overline{\Gamma}}$-$\mathrm{\overline{M}}$ path; 
the conduction band minimum (CBM) is found at the  $\mathrm{\overline{\Gamma}}$ point. 
The spin-polarized Dirac-cone surface states appear at the $\mathrm{\overline{\Gamma}}$ point, 
while the Dirac point is located only 0.05 eV above the Fermi energy at (\bs)$_{6}$ and the point is embedded into the valence bands at   (\bts)$_{6}$ and (\bt)$_{6}$. 
These electronic structures are apparently not useful for the spintronics applications since the carriers are easily doped into the bulk valence bands and hard to exploit the Dirac carriers. 
Even worse, the spin current may leak out from the surface to the bulk due to the metallic bulk band structure.
Therefore, it is needed to shift the Dirac point upward, away from the VBM to the middle of the energy gap. 
In order to achieve it,  we adopt the following strategy: (1) set four QLs of \bts\ at the inner (bulk) slab to open the wide bulk band gap and (2) load the surface material which has shallower electron potential onto bulk (\bts)$_{4}$ and make a hetero-structural junction so that the surface state energy is expected to be shifted upward. 
%
%A band structure of six-QLs $\mathrm{Bi_{2}Te_{2}Se}$ slab is shown in Fig. \ref{fig:band} (b). 
%As one can see, the spin-polarized Dirac-cone surface states (SS) appear at the $\mathrm{\overline{\Gamma}}$ point around $E_{F}$. 
%The valence band maximum (VBM) of the bulk layer ({\it i.e.,} inner four quintuple layers) appears at the $\mathrm{\overline{\Gamma}}$-$\mathrm{\overline{M}}$ path near the $\mathrm{\overline{\Gamma}}$ point, and an energy of the bulk VBM is -0.060eV below $E_{F}$. 
%The conduction band minimum (CBM) of the bulk layer (inner for quintuple layers) appears at the at the $\mathrm{\overline{\Gamma}}$ point, and an energy of the bulk CBM is 0.306eV above $E_{F}$. 
%Therefore, band gap of the bulk layer $E_{\rm{gap}}$ = 0.366eV. 
%The Dirac point which is the crossing point of the surface states is located at -0.095eV below $E_{F}$. 
%The Dirac point energy ($E_{\rm{DP}}$) evaluated from the bulk VBM is $E_{\rm{DP}}$ = -0.035eV. Bulk valence band crosses $E_{F}$, because Dirac point and surface states locate in bulk valence band. 
%In this thin film structure, spin current may leak out from surface to bulk because of the metallic bulk band structure.
%\par
The computational results of the band-gap and the Dirac cone in the homo-structural TI slabs are summerized in Fig. \ref{fig:cbvb}. 
%Here the work function $\phi$ is defined as energy difference between the vacuum level and the bulk VBM, 
%which is an indicator of the potential depth of each slab. 
In the following section, we will present results of hetero-structural slabs as combining two types of TIs. 
% which have different work functions. 
%Work function $\phi$ is defined as $E_{\rm{Vacuum \: Level}}$ - $E_{\rm{Bulk \: VBM}}$ in this paper. 
%These values are visualized in Fig. \ref{fig:cbvb}. 6-$\mathrm{Bi_{2}Te_{2}Se}$ has the large bulk band gap, but the Dirac point locates in bulk valence band so that $E_{\rm{DP}}$ = -0.035 $<$ 0. 
%Thus the bulk conduction is dominant in this structure, and it is difficult to observe the Dirac carrier which is the surface conduction near the Dirac point. 
%The bulk conduction is also dominant in 6-$\mathrm{Bi_{2}Te_{3}}$, because the Dirac point energy of 6-$\mathrm{Bi_{2}Te_{3}}$ $E_{\rm{DP}}$ = -0.035 is smaller than 0. 
%On the other hand, the surface conduction of the Dirac carrier may be observed in 6-$\mathrm{Bi_{2}Se_{3}}$ and 6-$\mathrm{Sb_{2}Te_{3}}$, because the Dirac point energy of 6-$\mathrm{Bi_{2}Se_{3}}$ is $E_{\rm{DP}}$ = +0.052 $>$ 0 and that of 6-$\mathrm{Sb_{2}Te_{3}}$ is $E_{\rm{DP}}$ = +0.101 $>$ 0. 
%Therefore the positions of the Dirac point energies in 6-$\mathrm{Bi_{2}Se_{3}}$ and 6-$\mathrm{Sb_{2}Te_{3}}$ are higher than the bulk valence band maximum energies of each slabs. 
%6-$\mathrm{Bi_{2}Se_{3}}$ and 6-$\mathrm{Sb_{2}Te_{3}}$ have a the desired band structure in ideal perfect crystals.

\subsection{${Bi_{2}Te_{2}Se}$-based heterostructural TI slab}

Following our strategy, several combinations of hetero-structural slabs by using \bts\ and other TIs were tested. 
Successful results were obtained only when \bt\ or \st\ was chosen for the surface QL. 
As shown in Fig. \ref{fig:band5} (d), the \btbts\ hetero-structural slab shows the ideal TI surface-band structure, in which the Dirac point is located at the middle of the band gap at the $\mathrm{\overline{\Gamma}}$ point. 
As compared with the band structure of (\bts)$_{6}$ (cfr. Fig. \ref{fig:band5}(c)), 
 the bulk band gap is kept large as 0.32 eV, while the Dirac point is shifted upward in energy by 0.19 eV. 
 The effect is simply caused by the replacement of outermost Se layer of the slab by Te layer.  

\begin{figure}
\center
\includegraphics[width=6cm]{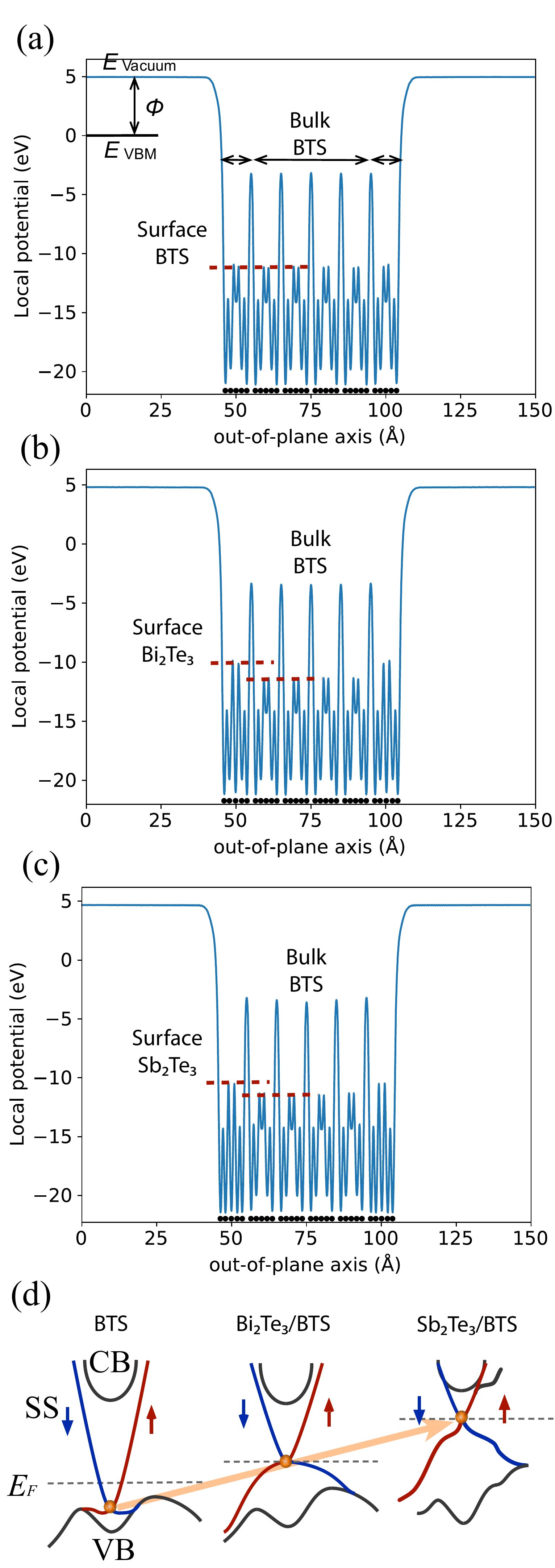}
\caption{Local potential of 
(a)  (\bts)$_{6}$, 
(b) \btbts, and % $\mathrm{Bi_{2}Te_{3}}$/4-$\mathrm{Bi_{2}Te_{2}Se}$/$\mathrm{Bi_{2}Te_{3}}$, and
(c) \sbbts\ slab % $\mathrm{Sb_{2}Te_{3}}$/4-$\mathrm{Bi_{2}Te_{2}Se}$/$\mathrm{Sb_{2}Te_{3}}$  slab 
plotted along the out-of-plane direction. 
The local potential energy is defined as $V_{\rm{local}}$ = $V_{\rm{ionic}}$ + $V_{\rm{Hartree}}$ with respect to the  $E_{\rm F}$ as the origin of energy. Black dots at the graph bottom  indicate the atomic position. 
$E_{\rm Vacuum}$, $E_{\rm VBM}$, and $\phi$ denote the vacuum energy, the valence energy maximum, and the work function of the slab, respectively. 
(d) Schematic picture of surface-band states of those slabs. 
}
\label{fig:locpot}
\end{figure}

In order to understand the shift of the surface-state energy, we evaluated the work function of slab structures. 
Table \ref{tbl:DP} (upper part) shows the work function $\phi$ of 6QL slabs that contain a single TI material, where the work function is defined as energy difference between bulk VBM and the potential at the vacuum region between slabs. 
That can be an indicator of the potential depth of each slab. 
It can be seen that the work function at (\bt)$_{6}$, $\phi$=4.95eV, 
 is slightly shallower than that in  (\bts)$_{6}$, $\phi$=5.02eV. 
The difference of these work functions, $\Delta\phi$ = +0.07 eV, is indeed responsible for the upward shift  of the Dirac point in  \btbts\ multiQL; 
the surface bands arising from the outmost \bt\ QL is shifted up relatively to the bulk bands arising from the inner (\bts)$_{4}$ QLs. 
%By comparing the surface bandstructures of (\bts)$_{6}$  and \btbts, it can be seen that the Dirac point is shifted upward by 0.07 eV (cfr. Fig.\ref{fig:band5}(c)  and Fig.\ref{fig:band5} (d)). 
%kyama 2019 Nov.
In the case of a multiQL $\mathrm{Sb_{2}Te_{3}}$/$(\mathrm{Bi_{2}Te_{2}Se})_{4}$/$\mathrm{Sb_{2}Te_{3}}$, as shown in Fig. \ref{fig:band5} (e), the Dirac point is shifted further upward by $\sim$0.003 eV and approaches  the bulk CBM. 
The trend is also depicted in Fig.\ref{fig:locpot} (d). 
%The bulk valence band maximum locates -0.160eV below $E_{F}$ at the $\mathrm{\overline{\Gamma}}$-$\mathrm{\overline{M}}$ path near the $\mathrm{\overline{\Gamma}}$ point. 
%The bulk conduction band minimum locates +0.105eV above $E_{F}$ at the $\mathrm{\overline{\Gamma}}$ point. 
%Therefore the bulk band gap energy is $E_{\rm{gap}}$ = 0.265eV. 
%Position of the Dirac point ($E_{\rm{DP}}$) relative to the bulk valence band maximum is $E_{\rm{DP}}$ =  +0.158eV. 
%With respect to the (\bts)$_{6}$, the Dirac point shifts upward +0.193eV in energy by joining $\mathrm{Sb_{2}Te_{3}}$ layer to the surface of 4-$\mathrm{Bi_{2}Te_{2}Se}$. 
%Work function of $\mathrm{Sb_{2}Te_{3}}$ which is surface layer is 4.650eV. 
%Work function of $\mathrm{Sb_{2}Te_{3}}$ is smaller than that of $\mathrm{Bi_{2}Te_{3}}$. 
%Therefore, the Dirac point shifts upward more in $\mathrm{Sb_{2}Te_{3}}$/4-$\mathrm{Bi_{2}Te_{2}Se}$/$\mathrm{Sb_{2}Te_{3}}$ than in $\mathrm{Bi_{2}Te_{3}}$/4-$\mathrm{Bi_{2}Te_{2}Se}$/$\mathrm{Bi_{2}Te_{3}}$.
This can be understood by calculating the work function at (\st)$_{6}$, $\phi$=4.65eV, that is smaller than those at (\bt)$_{6}$ and (\bts)$_{6}$. 
%Local potential reflects the potential around the electron of valence band maximum. 
%In fact, local potential data of this structure (Fig. \ref{fig:locpot} (b)) indicates that local potential energy of surface quintuple layer which is $\mathrm{Bi_{2}Te_{3}}$ is higher than bulk four quintuple layers which are 4-$\mathrm{Bi_{2}Te_{2}Se}$. 
%The local potential energy difference between the surface layers and the bulk layers depends on the work functions. 
%Thus the Dirac point shifts upward. 
%
The energy difference caused between the surface QL and the inner QLs can be visualized by plotting  
the electron local potential. % that electron feels. 
Figure \ref{fig:locpot} (a)-(c) shows the local potential along the out-of-plane axis in a super cell containing 6QL slab and the vacuum region. 
It is seen that the local potential is rather flat in \bts\ slab region and becomes shallower at a surface QL 
of \bt\ and \st. 
% with respect to the inner QLs in \btbts\ and \sbbts\ slabs.  

\begin{figure*}[hbt!]
\center
\includegraphics[width=17cm]{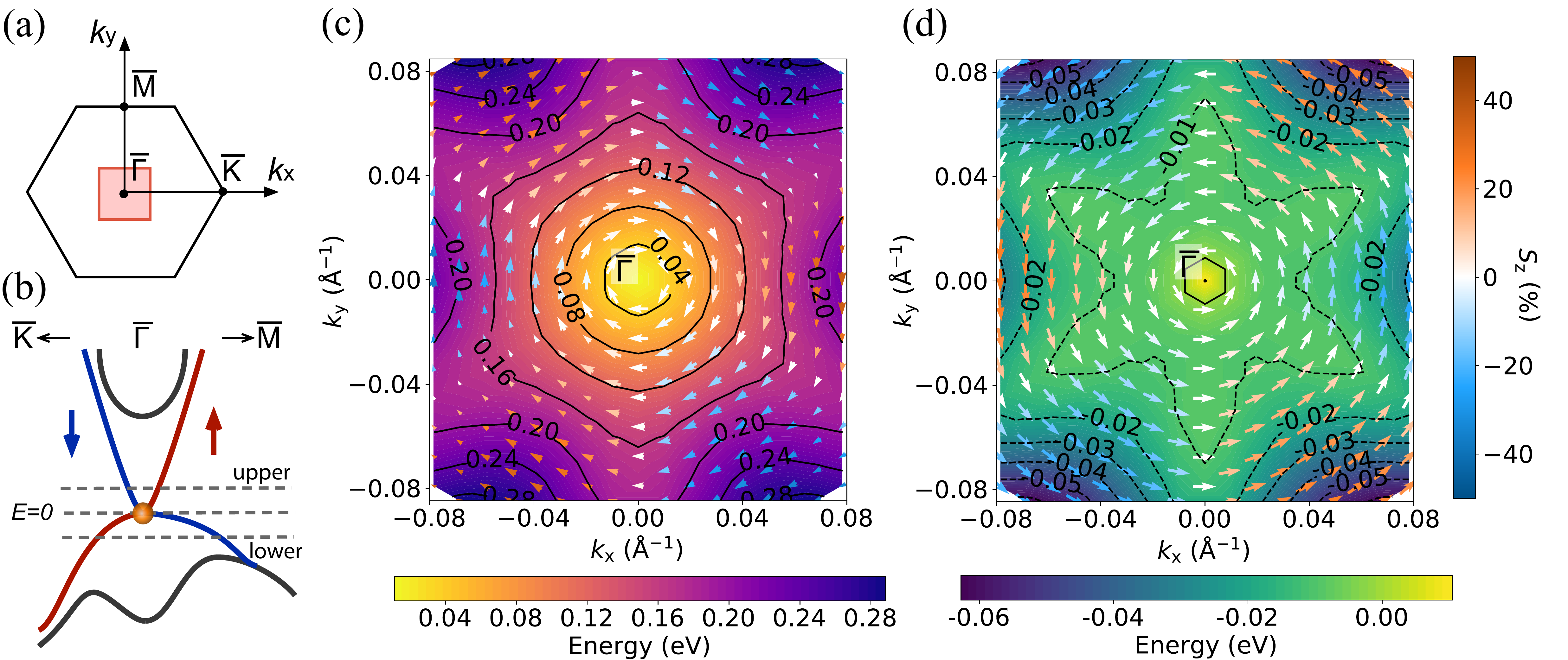}
\caption{
(a) The hexagonal Brillouin zone of \btbts. 
The red square shows the plotting region of spin texture. % at panel (c) and (d). 
(b) The schematic band structure of upper and lower Dirac-surface states. 
Calculated spin texture %of the Dirac surface states in the \btbts\ multilayer slab 
for (c) the upper Dirac surface band  and for  (d) the lower Dirac surface band.
Lengths and directions of arrows indicate in-plane ($S_{\rm{x}}$, $S_{\rm{y}}$) spin components. 
Colors of arrows indicate out-of-plane spin $S_{\rm{z}}$ component. 
Contours show eigenvalues of Dirac surface bands in each ${k}$-points. 
Energy is measured with respect to the $E_{\rm F}$. 
}
\label{fig:spin_BTS_T}
\end{figure*}

The Dirac surface state is characterized by intriguing properties such as %the surface  spin polarization and the
 spin-momentum locking caused by  lack of space-inversion symmetry and presence of spin-orbit coupling. 
The isocontour of the surface-band energy and the spin texture of the Dirac surface states in \btbts\  slab are shown in Fig. \ref{fig:spin_BTS_T}. 
It is noted that the constant-energy contour ofthe lower Dirac band shows  strongly warped shape (that looks like a snowflake) while the upper Dirac band shows hexagonal-looking energy contour. 
The difference may come from the different characters of the bands; the upper Dirac band consists of mainly Bi $p$ orbital state and the lower consists of mainly Te $p$ state. 
The warping effect can be also seen in \bt\ surface\cite{warp}, while the effect is strongly enhanced in \btbts, where 
the outmost \bt\ surface QL is located between the vacuum region and \bts\ QLs and create the rather non-centrosymmetric electronic potential at the surface Bi atom. 
According to a $k\cdot p $ perturbation theory \cite{warp} with spin-orbit coupling with  $\mathrm C_{3v}$-symmetry 
 around the $\Gamma$ point,  %Brillouin zone,
effective Hamiltonian up to the third order of $\vec{k}$ reads 
\begin{equation}
H(\vec{k})=E_{0}(\vec{k}) + \varv_{k}\left(k_{x}\sigma_{y}-k_{y}\sigma_{x}\right)+\frac{\lambda}{2}\left(k_{+}^{3}+k_{-}^{3}\right)\sigma_{z}.
\end{equation} 
Here, the first term $E_{0}(\vec{k})=k^{2}/(2m^{*})$ is free-electron dispersion with the mass $m^{*}$. 
The second term with the Dirac velocity $\varv_{k}=\varv \left(1+\alpha k^{2}\right)$ and the in-plane spin momentum $\sigma_{x, y}$ 
is responsible for two-dimensional Rashba effect. 
The last term is essential for 
the Fermi-line warping (here $k_{\pm}= k_{x}\pm i k_{y}$) and  also  emergence of the out-of-plane spin polarization $\sigma_{z}$. 
In fact, as shown in Fig. \ref{fig:spin_BTS_T} (c), 
the spin texture of  the lower Dirac band shows not only the 
Rashba-type tangential spin vortex but  also the out-of-plane spin component $S_{z}$ along the $\bar\Gamma - \bar{\mathrm K}$ line away from the $\bar \Gamma$ point.\cite{sz} 
The upper and lower bands basically show the opposite directions of spin polarization. 
%, and the out-of-plane spin component on the surface states appear. 
%It is caused by $k^3$ term of spin-orbit coupling\cite{warp} at the rhombohedral crystal systems.
%Its term is enhanced by inversion symmetry broken at the interface of $\mathrm{Bi_{2}Te_{3}}$/4-$\mathrm{Bi_{2}Te_{2}Se}$ junction

\begin{table}
\caption{
%In-plane lattice constants $a$ taken from experimental studies, 
 Calculated band gap of the {\it bulk} state $E_{\rm{gap}}$, 
 work function  defined as $\phi=E_{\rm{Vacuum \: Level}}$ - $E_{\rm{Bulk \: VBM}}$, and
   Dirac point energy ($E_{\rm{DP}}$) with respect to the {\it bulk} VBM %$E_{\rm{Bulk \: VBM}}$
 of the 6QL slabs. 
Here the {\it bulk} state is regarded as the state of bands contributed from the inner several QLs in each slab.
Position of Dirac point ($E_{\rm{DP}}$) in each topological insulators multilayers (Energy from bulk valence band maximum $E_{\rm{Bulk \: VBM}}$). 
$a$ is a in-plane lattice constant of each multilayers. 
$E_{\rm{gap}}$ is a band gap of the bulk states. 
The bulk states is regarded as the contribution of the bulk layers to the band structure. 
The bulk layers are inner layers of slabs. 
Work function $\phi$ is defined as $E_{\rm{Vacuum \: Level}}$ - $E_{\rm{Bulk \: VBM}}$ here. 
BTS stands for \bts. 
The ideal $E_{\rm{DP}}$ values (i.e., the Dirac cone appears in the middle of the wide energy gap) are highlighted. 
}
\label{tbl:DP}
\begin{center}

\begin{tabular}{crrrr}
\hline
\hline
%\multicolumn{1}{c}{Slab} & \multicolumn{1}{c}{$a$ ($\AA$)} & \multicolumn{1}{c}{$E_{\rm{gap}}$ (eV)} & \multicolumn{1}{c}{$\phi$ (eV)} & \multicolumn{1}{c}{$E_{\rm{DP}}$ (eV)}  \\
\multicolumn{1}{c}{Slab} &  \multicolumn{1}{c}{$E_{\rm{gap}}$} & \multicolumn{1}{c}{$\phi$} & \multicolumn{1}{c}{$E_{\rm{DP}}$}  \\
\multicolumn{1}{c}{} &  \multicolumn{1}{c}{(eV)} & \multicolumn{1}{c}{(eV)} & \multicolumn{1}{c}{(eV)} \\

\hline
$\mathrm{(Bi_{2}Se_{3})_{6}}$ &  0.305 & 5.603 & +0.052  \\
$\mathrm{(Bi_{2}Te_{3})_{6}}$ & 0.251 & 4.946  & -0.121  \\ %kosaka value -0.035  \\
$\mathrm{(BTS)_{6}}$ & 0.366 & 5.024  & -0.091 \\   %kosaka value -0.035  \\
$\mathrm{(Sb_{2}Se_{3})_{6}}$ & 0.314 & 5.404 &  None \\
$\mathrm{(Sb_{2}Te_{3})_{6}}$ & 0.267 & 4.650 & \textbf{ +0.101 } \\
\hline
%\end{tabular}

%\begin{tabular}{crrrr}
%\hline
%\multicolumn{1}{c}{Multilayer Structure} & \multicolumn{1}{c}{$a$} & \multicolumn{1}{c}{$E_{\rm{gap}}$} & \multicolumn{1}{c}{$\phi$} & \multicolumn{1}{c}{$E_{\rm{DP}}$} \\
%\multicolumn{1}{c}{} & \multicolumn{1}{c}{($\AA$)} & \multicolumn{1}{c}{(eV)} & \multicolumn{1}{c}{(eV)} & \multicolumn{1}{c}{(eV)} \\
%\hline
$\mathrm{Sb_{2}Se_{3}}$/$(\mathrm{Bi_{2}Se_{3}})_{4}$/$\mathrm{Sb_{2}Se_{3}}$ &  0.269 & 5.471 & +0.016 \\
$\mathrm{Bi_{2}Te_{3}}$/$(\mathrm{Bi_{2}Se_{3}})_{4}$/$\mathrm{Bi_{2}Te_{3}}$ & & Metal &  \\
$(\mathrm{Bi_{2}Te_{3}})_{2}$/$(\mathrm{Bi_{2}Se_{3}})_{4}$/$(\mathrm{Bi_{2}Te_{3}})_{2}$ & & Metal &  \\
\hline
$(\mathrm{Sb_{2}Te_{3}})_2$/$(\mathrm{Bi_{2}Te_{3}})_2$/$(\mathrm{Sb_{2}Te_{3}})_2$ &  0.199 & 4.940  & +0.035  \\
$\mathrm{Sb_{2}Te_{3}}$/$(\mathrm{Bi_{2}Te_{3}})_4$/$\mathrm{Sb_{2}Te_{3}}$ &  0.237 & 4.949 & +0.006 \\
$(\mathrm{Sb_{2}Te_{3}})_2$/$(\mathrm{Bi_{2}Te_{3}})_4$/$(\mathrm{Sb_{2}Te_{3}})_2$ &  0.165 & 4.891 & -0.031 \\
$\mathrm{Bi_{2}Se_{3}}$/$(\mathrm{Bi_{2}Te_{3}})_{4}$/$\mathrm{Bi_{2}Se_{3}}$ & & Metal &  \\
$(\mathrm{Bi_{2}Se_{3}})_{2}$/$(\mathrm{Bi_{2}Te_{3}})_{4}$/$(\mathrm{Bi_{2}Se_{3}})_{2}$ & & Metal &  \\
\hline
$\mathrm{Bi_{2}Se_{3}}$/$(\mathrm{BTS})_{4}$/$\mathrm{Bi_{2}Se_{3}}$ &   & Metal &  \\
$\mathrm{Bi_{2}Te_{3}}$/$(\mathrm{BTS})_{4}$/$\mathrm{Bi_{2}Te_{3}}$ &  0.321 & 4.947 & \textbf{+0.096} \\ %kosaka +0.140\\
$\mathrm{Bi_{2}Te_{3}}$/$(\mathrm{BTS})_{6}$/$\mathrm{Bi_{2}Te_{3}}$ &  0.290 & 4.923  &\textbf{ +0.113}\\
$(\mathrm{Bi_{2}Te_{3}})_{2}$/$(\mathrm{BTS})_{2}$/$(\mathrm{Bi_{2}Te_{3}})_{2}$ &  0.253 & 4.853  & +0.051\\
$(\mathrm{Bi_{2}Te_{3}})_{2}$/$(\mathrm{BTS})_{4}$/$(\mathrm{Bi_{2}Te_{3}})_{2}$ &  0.221 & 4.817  & +0.029\\
$(\mathrm{Bi_{2}Te_{3}})_2$/$(\mathrm{BTS})_6$/$(\mathrm{Bi_{2}Te_{3}})_2$ &  0.172 & 4.825  & +0.038 \\
$(\mathrm{Bi_{2}Te_{3}})_3$/$(\mathrm{BTS})_4$/$(\mathrm{Bi_{2}Te_{3}})_3$ &  0.160 & 4.782  & -0.023 \\
$(\mathrm{Bi_{2}Te_{3}})_3$/$(\mathrm{BTS})_6$/$(\mathrm{Bi_{2}Te_{3}})_3$ &  0.143 & 4.806 & -0.023  \\
$\mathrm{Sb_{2}Se_{3}}$/$(\mathrm{BTS})_4$/$\mathrm{Sb_{2}Se_{3}}$ &  & Metal &  \\
$\mathrm{Sb_{2}Te_{3}}$/$(\mathrm{BTS})_4$/$\mathrm{Sb_{2}Te_{3}}$ &  0.265 & 4.836 & \textbf{ +0.158 }\\
$\mathrm{Sb_{2}Te_{3}}$/$(\mathrm{BTS})_6$/$\mathrm{Sb_{2}Te_{3}}$ &  0.263 & 4.830  & \textbf{+0.146} \\
$(\mathrm{Sb_{2}Te_{3}})_2$/$(\mathrm{BTS})_2$/$(\mathrm{Sb_{2}Te_{3}})_2$ &  0.225 & 4.761 & +0.150  \\
$(\mathrm{Sb_{2}Te_{3}})_2$/$(\mathrm{BTS})_4$/$(\mathrm{Sb_{2}Te_{3}})_2$ &  0.189 & 4.724 & +0.127 \\
$(\mathrm{Sb_{2}Te_{3}})_2$/$(\mathrm{BTS})_6$/$(\mathrm{Sb_{2}Te_{3}})_2$ &  0.229 & 4.664 & \textbf{+0.089} \\
$(\mathrm{Sb_{2}Te_{3}})_3$/$(\mathrm{BTS})_4$/$(\mathrm{Sb_{2}Te_{3}})_3$ &  0.136 & 4.651 & +0.050 \\
$(\mathrm{Sb_{2}Te_{3}})_3$/$(\mathrm{BTS})_6$/$(\mathrm{Sb_{2}Te_{3}})_3$ &  0.207 & 4.569 & -0.005 \\
\hline
$\mathrm{Bi_{2}Te_{3}}$/$(\mathrm{Sb_{2}Te_{3}})_2$/$\mathrm{Bi_{2}Te_{3}}$ & 0.271 & 4.685 & +0.007  \\
\hline
\hline
\end{tabular}
\end{center}
\end{table}

%\subsection{Multilayers consist of two kinds of topological insulator materials}

In order to find the optimal Dirac surface state, more than 20 types of  hetero-structural topological QL slabs were examined. 
 The calculated results of the energy gap and the Dirac-point energy are summarized  in Table \ref{tbl:DP}. 
%Calculation results of the each topological insulator multilayers are sumerized. 
%Table \ref{tbl:DP} shows that the Dirac point can be optimized by tuning work functions between the surface layers and the inner layers. 
Among these heterostructure candidates, the ideal Dirac surface state which form the crossing point at the middle of the wide band gap is found at 
$\mathrm{Bi_{2}Te_{3}}$/$\mathrm{(Bi_{2}Te_{2}Se)_{4}}$/$\mathrm{Bi_{2}Te_{3}}$, 
$\mathrm{Bi_{2}Te_{3}}$/$\mathrm{(Bi_{2}Te_{2}Se)_{6}}$/$\mathrm{Bi_{2}Te_{3}}$, 
$\mathrm{Sb_{2}Te_{3}}$/$\mathrm{(Bi_{2}Te_{2}Se)_4}$/$\mathrm{Sb_{2}Te_{3}}$, 
$\mathrm{Sb_{2}Te_{3}}$/$\mathrm{(Bi_{2}Te_{2}Se)_6}$/$\mathrm{Sb_{2}Te_{3}}$, and 
$(\mathrm{Sb_{2}Te_{3}})_2$/$\mathrm{(Bi_{2}Te_{2}Se)_6}$/$(\mathrm{Sb_{2}Te_{3}})_2$.  
In these slabs, %, as well as $\mathrm{Bi_{2}Te_{3}}$/$(\mathrm{BTS})_{4}$/$\mathrm{Bi_{2}Te_{3}}$ 
as discussed above, 
the bulk \bts\ QLs opens the wide band gap ($E_{\rm gap}\sim 0.3$ eV) and the surface Dirac cone originating from \bt\ and \st\ QLs appears at the middle of the gap; 
  the shallower work function of the surface QL with respect to that of the bulk QL 
 slightly shifts up the surface state energy; 
e.g., $(\mathrm{Sb_{2}Te_{3}})_6$ slab shows $\phi=4.95$ eV and $\mathrm{(Bi_{2}Te_{2}Se})_6$ shows $\phi=5.02$ eV. 
As varying the number of  \st\ and \bts\ QLs at
$(\mathrm{Sb_{2}Te_{3}})_n$/$\mathrm{(Bi_{2}Te_{2}Se)_m}$/$(\mathrm{Sb_{2}Te_{3}})_n$ slabs, 
 it is feasible to tune the position of the Dirac cone in the wide energy region ($0<E_{\rm DP}<0.16$). 
On the other hand, when we combine  two types of QL-materials whose work functions are much different, the hetero-structural slab goes metallic. 
For example, the \bs\ and \bt\ 6QL slabs show $\phi=$ 5.60 and 4.95 eV, respectively, 
and then the hetero-structural combination of these QLs is found to be metallic. 
In $(\mathrm{Bi_{2}Te_{3}})_n$/$\mathrm{(Bi_{2}Se_{3})_m}$/$(\mathrm{Bi_{2}Te_{3}})_n$ slabs, 
it is metallic not only at the surface state but also at the bulk state because the 
CBM of \bs\ overlaps with the VBM of \bt\ (see Fig. \ref{fig:cbvb}). 

%Multilayers consist of two kinds of materials whose work functions mush vary in energy become metallic band structure, because the valence band of one material overlaps with the conduction band of the other material. 
%As example, $\it{n}$-$\mathrm{Bi_{2}Se_{3}}$/$\it{m}$-$\mathrm{Bi_{2}Te_{3}}$/$\it{n}$-$\mathrm{Bi_{2}Se_{3}}$ become metal not only surface states but also bulk states, because the valence band of $\mathrm{Bi_{2}Te_{3}}$ overlaps with the conduction band of $\mathrm{Bi_{2}Se_{3}}$ (see Fig. \ref{fig:cbvb}).

\section{Summary}

First-principles DFT calculations  have been carried out on the hetero-structural multi-QL slabs of topological insulators. 
We have demonstrated that 
 the Dirac-point position can be shifted to the middle of the band gap by combining the appropriate pair of QLs of different materials. 
The surface-band energy can be further tuned by varying the number of QLs in the slab. 
Since the controllability of Dirac cone can be acquired   by simple combination of the conventional topological insulators, 
it may be easily confirmed by following experimental studies and promising for future spintronics applications.

\acknowledgment

KY acknowledges Takafumi Sato and Seigo Souma for the fruitful discussions on topological materials. 
This work was supported by JSPS Kakenhi (No. 17H02916 and 18H04227) and by JST-CREST (No: JPMJCR18T1). 
A part of the computation in this work has been done by using the facilities of the Supercomputer Center, the Institute for Solid State Physics, the University of Tokyo.
The crystallographic figure was generated using VESTA program.\cite{vesta}

%\appendix
%\section{}

%thanks.

%\profile{Takao Kosaka}{was born in Hyogo, Japan in 1995. ...}

%Unused bibitems


\begin{thebibliography}{99}
\bibitem{lat.sb2se3} Since the experimental lattice constants for \sbs\ were not found in literatures, the lattice structure was obtained by lattice optimization.
\end{thebibliography}
\end{document}